# Symmetry and Contextuality

Hartmann Römer

Physikalisches Institut der Universität Freiburg
Hermann-Herder-Str. 3, 79104 Freiburg
hartmann.roemer@physik.uni-freiburg.de
http://omnibus.uni-freiburg.de/~hr357

# Abstract

The general concept of symmetry is realized in manifold ways in different realms of reality, such as plants, animals, minerals, mathematical objects or human artefacts in literature, fine arts and society. In order to arrive at a common ground for this variedness a very general conceptualization of symmetry is proposed:
> *Existence of substitutions, which, in the given context, do not lead to an essential change.*

This simple definition has multiple consequences:
-The context dependence of the notion of symmetry is evident in the humanities but by no means irrelevant yet often neglected in science. The subtle problematic of concept formation and the ontological status of similarities opens up.
-In general, the substitutions underlying the concept of symmetry are not really performed but remain in a state of virtuality. Counterfactuality, freedom and creativity come into focus. The detection of previously hidden symmetries may provide deep and surprising insights.
-Related to this, due attention is devoted to the aesthetic dimension of symmetry and the breaking of it.
-Finally, we point out to what extent life is based on the interplay between order and freedom, between full and broken symmetry.

# Introduction

At first sight, symmetry looks like a rather unproblematic concept for most people. As paradigms we usually have a stock of interpreted elements of reality in our mind for which a symmetry of some kind seems to be evident. Some of these examples are almost universal and shared by virtually every one, others may arise from personal or professional experience. Physicists, sculptors, poets and musicians may and will conceive symmetries in different ways, and realising a previously unnoticed symmetry may be felt as a revelation, sometimes granting an insight into deep hidden layers of reality.

These simple observations prompt at least two fundamental questions:
1. What do the symmetries perceived by different people under different points of view have in common? Is it possible to arrive at a definition of symmetries general enough to comprise these various notions without missing what is felt to be the core of the concept of symmetry?
2. What is the ontological status of symmetries? Are they mere inventions of the human mind as their subject dependence seems to suggest or are they full-fledged elements of reality as suggested by their widespread intersubjectivity and their indispensable helpful function for ordering, interpreting and understanding our perceptions in our world?

After these introductory remarks we shall proceed as follows:

First, we shall give examples of both overt and covert symmetries in different fields like zoology, botany, crystallography, mathematics, physics and fine arts and human artefacts. Guided by these examples we propose a general definition of the concept of symmetry as an answer to the first question. Seriality will be identified as a special case of symmetry. Thereafter we investigate the consequences of our definition. The relatedness of symmetry with the processes of concept formation and idealisation will come into focus, leading to a tentative answer to the second question. Moreover, we shall see how closely symmetry and symmetry breaking are akin to aesthetics, freedom, creativity and liveliness.

# Miscellaneous Examples of Symmetries

(1) Animals: Any zoology textbook tells us that all higher animals, including mammals, reptiles, fishes, insects and crabs, are "bilateria", i.e. they show an approximate left-right symmetry. Two sides of their bodies are mirror images of one another with respect to a plane transecting their body from top to bottom. (Even snails and starfishes are only apparent exception as revealed by their larvae stadium.) A further more blurred symmetry is "metamery", a tendency of repetition in longitudinal direction, visual, for instance in the repetitive nature of the rips and the vertebral spine.

(2) Plants: Flowers sometimes also exhibit mirror symmetry (e.g. the snapdragon) but more often symmetries under discrete rotations. Two-fold, three-fold, four-fold, five-fold and six-fold rotational symmetries can be found. Beyond six, seven- or eight-fold symmetries are not observed but only indeterminate large numbers occur (e.g. the daisy). The position of leaves around the stem often shows a screw symmetry under a rotation by an integer fraction of 360 degrees combined with a translation along the axis by a certain fixed amount. A large fluorescence often possesses a spiral symmetry combining a little rotation with a displacement in radial direction. For instance, the seeds of the sunflower are arranged in a spiral pattern. Here we have a first example of a covert and surprising regularity. The numbers $f_n$ of seeds in the n-th layer of the spiral very closely follow the recursion relation $f_n = f_{n-1} + f_{n-2}$ of the famous Fibonacci sequence. Remarkably, the ratios $f_n/f_{n+1}$ tend towards the golden ratio. This assures saving space: as much as possible the seeds of the following layers come to lie in the gaps left by the preceding layer.

(3) Crystals: The building blocks of a crystal are arranged in an extremely regular pattern. They are placed on a periodic grid generated by shifts in three different directions by a fixed amount for every direction. In addition to these translations, a crystal may be symmetric under rotations, mirror reflections and combinations of these elements. A surprising instance of "a priori physics" emerges from this [1]: there are exactly 14 possible types of translational grids, symmetry rotations are possible only for angles of 180, 120, 90 or 60 degrees and can be combined in exactly 32 different ways; and altogether there are precisely 230 different types of crystals, all of which being realised in nature. (More recently, so-called quasicrystals [2] have been found with, e.g., five- or seven-fold rotational symmetries. But for them the requirement of periodicity is weakened to quasi-periodicity.)

(4) Mathematics: A cube is a good example of a highly symmetric geometric mathematical object. It has eight corners, twelve edges and six faces. It is transformed into itself by many symmetry operations also called symmetry transformations: mirroring about planes through the centre and parallel to a face, threefold rotations about a corner, twofold rotations about the midpoint of an edge and fourfold rotations about the centre of a face with respect to an axis through the centre of the cube. These symmetries are exact and not only approximate. In fact, exact, *unbroken* symmetries are only possible for idealised entities like mathematical objects. The cube is just the most well-known one of the *platonic bodies,* all of whose faces must be regular polytopes of the same type with an equal number of them meeting at each corner. There are exactly five platonic bodies: the tetrahedron, the octahedron and the icosahedron with three, four and five equilateral triangles meeting at each corner, the cube with three squares at each corner and the dodecahedron with three regular pentagons meeting at each corner. Remarkably, the shape of the platonic bodies is completely determined by the symmetry operations which they allow.

As an arithmetic example of mathematical symmetry we should like to mention the *Mandelbrot set* [3]. Its definition is easy. Consider the following operation on complex numbers:
$$z_{n+1} = z_n^2 + c \;.$$
Start from $z_0 = 0$ and apply the operation again and again. The Mandelstam set is the set of those numbers c, for which the iterated images of $z_0$ do not escape to infinity. It is, of course, invariant under the above operation. At its boundary and a little bit beyond of it, the Mandelstam set shows a structure of inexhaustible complexity. Zooming into smaller and smaller scales reveals new structures emerging for the first time, just to reappear later after increased magnification. In this sense, the Mandelstam set is *self-similar* with zooming in as a symmetry operation. I think, one can hardly avoid the uncomfortable and eerie impression of falling through a trap door to find oneself in a world, which is strongly regulated by a simple principle, yet unpredictable and in no way made for man, although the emerging structures, sometimes in a tantalising way, are reminiscent of frost patterns, fronds of fern, seahorse tails or eyes on a peacock tail. Here we see how the detection and investigation of a symmetry may lead into unfathomable depths.

(5) Laws of physics: The fundamental laws of physics are equations such that admissible physical states or processes must be a solution of these equations. Because all points and directions in space as well as all time points are physically equivalent, every state or process arising from an admissible state or process by a displacement in space or time or by a rotation must also be admissible. These symmetry properties of the fundamental laws of physics have a deep and surprising consequence called *Noether's theorem* [4]:

Energy conservation is a consequence the symmetry under time translations and the conservation of momentum and angular momentum are due to symmetry under spatial translations and rotations. In contemporary physics also more abstract symmetry operations different from space-time transformations (for example gauge transformations) are considered as a powerful tool to find fundamental physical laws, which are strongly constrained by their symmetries just as spatial symmetries determine the shapes of the Platonic bodies.

Notice that the solutions of the fundamental physical equations are in general less symmetric than the equations themselves. Rotational symmetry of the fundamental equations, of course, does not mean, that every physical object must be round. It only tells us that rotating an admissible physical configuration produces another admissible configuration with the same physical properties. In general, admissible states and processes will break the high symmetry of the physical laws.

(6) Human artefacts: In the previous examples, we saw that symmetry relations are not always as straight-forward as one might expect, but may be hidden and full of surprises. For example, the symmetries of the fundamental laws of nature are still a main subject of active research. Entering the realm of human artefacts, a brief glimpse already suffices to convince us that the manifold of different symmetry-like relations, which can be perceived and identified, is vast and in fact inexhaustible.

- Handicraft produces structures like ornaments, tessellations and mandalas with an enormous variety of (approximate) spatial symmetries.
- This is also true in visual arts and architecture, which exhibit unlimited possible similarity or contrast correspondences of colours and shapes.
- In poetry, symmetry-like correspondences are realised in rhymes, assonances or strophe patterns. They appear in both syntactic and semantic correspondences like the *parallelismus membrorum* in the psalms, where a motive is presented twice in a modified way in two successive verses. They also appear in more abstract, subtle correspondences and contrasts of various elements and the way they are expressed, as well as in a work's overall structure.
- Music is full of correspondences such as harmonies and disharmonies, repetitions, "leitmotivs" or counterpoint and other symmetry relationships between different voices. Rhythm as a very particular time structure of continuous but slightly varied repetition is a constitutive symmetry element in music.
- Even in reflective disciplines like philosophy, symmetry elements can be found in the architecture of the building of thoughts. They can also become topical in the presentation and development of ideas.

There is an infinite manifoldness of possible similarities, contrasts, variations, inversions and allusions. In most of these cases, the process of identifying a symmetry and the way of its realisation is far from trivial. Moreover, almost universally, symmetry has an aesthetic connotation here.

## Attempt at a General Definition of Symmetry and its Consequences

Trying to find a least common denominator for all the examples given above, we can rely on the following observations: There were always substitutions replacing parts of the system by other parts such that the system did not change very much. For instance, for bilateral symmetry the substitution consisted in an exchange of the left and right side by mirroring. For rotational symmetries the substitutions were rotations transforming parts of the system into one another. Other possible substitutions are shifts in space or time. One easily finds examples for which space or time substitutions are accompanied by other changes. Even inversions will occur as can be seen by the example of colour. In such cases the symmetry has contrastive features.

The fundamental symmetry substitution of the Mandelstam set is the replacement of z by $z^2 + c$. The "zoom in" symmetry operations of the Mandelstam set result from it.

In particular, for human artefacts the substitutions were of a very general and sometimes abstract nature, such as replacing a word by a rhyming one or a motive by a similar one. It is in general not clear from the outset what "similarity" means in every instance. What are the substitutions that "do not change very much", but are considered admissible because they leave those features of the system intact that are considered essential? Such decisions depend on a given *context,* under which the system is looked at. A general definition of symmetry taking all this into account could be formulated as follows:

*Symmetry of a system means existence of substitutions within it, which,* **in a given context**, *do not lead to an essential change.*

The discussion of this context dependence of the "change without change" concept of symmetry will be one of our main concerns. Before enlarging upon the crucial importance of contextuality, I shall introduce some additional concepts related to symmetry:

The term *seriality* refers to the special case in which symmetry substitutions involve repetition of elements in space or time.

Symmetry substitutions that do not lead to an essential change may be repeated or performed one after another to generate other symmetry substitutions. The resulting mathematical structure is called a *semigroup* of symmetry substitutions. Instead of symmetry substitutions we will synonymously use the terms *symmetry operations* or *symmetry transformations.* For formal closure, it will be appropriate to introduce a trivial *unity transformation,* which, by definition, simply substitutes the elements by themselves thereby not leading to any change. Moreover, in many situations it may be possible to undo or reverse a symmetry operation, such that applying a reversed operation after an operation leads to the unity transformation. If this is possible for every symmetry operation, the corresponding structure is a *group* of symmetry transformations. Rotations and translations are examples of reversible transformations, whereas the transformations of the Mandelstam set are not reversible.

Context dependence is a salient feature of our definition of symmetry. It may come as a surprise for someone having mainly examples from botany or architecture in mind since the context seems to be naturally given and too trivial to be a worthy subject for attention and reflexion. On closer consideration, however, it becomes evident that even in such realms this is not entirely clear. First of all, symmetries need not be so cogent that nobody can overlook them. The sunflower gave an example of a covert symmetry, whose detection is a real discovery. Finding the symmetries of fundamental physical laws is a subtle and highly creative task still at the cutting edge of research. On the other hand, no one can be forced to see a symmetry, even if it seems to be evident for other observers. In particular, symmetries of natural objects are only approximate and never exact, what raises the question why we see them at all. For human artefacts the contextuality of symmetries is evident and indispensable if this concept shall somehow remain applicable in a meaningful way at all. In other fields, the contextuality of symmetries may be an important, frequently overlooked, lesson.

In general, the similarities underlying a symmetry are neither trivially read off from an object nor are they conceived as completely free and arbitrary inventions of the human mind. The process of identifying or constituting similarities is in fact encumbered with all the fundamental philosophical problems of the formation and the ontological status of concepts. Concept formation will always be related to the formation of an "*equivalence class*" uniting a plenitude of "similar" objects in into one conceptual unity. Exact symmetries can only occur between such idealised objects like equivalence classes, in which, under the given context, all of its members are identified. This can only be true for mathematical symmetries. Without idealisation, the elements of an equivalence class are still kept apart and the perfect symmetry of the idealisation appears slightly or even strongly broken. In short, principles are more symmetric than their realisations.

As far as the ontological status of symmetries or, more general, concepts is concerned, the philosophical dispute is a pervasive subject matter of philosophy. In medieval times the problem of *universals* was the subject of heated discussions*: realism* accepted them as objective entities independent of the human mind*, nominalism* attributed to them a lower ontological status as mere names. The dispute was never resolved and lives on until today under different disguises, for instance as the contrast between *essentialism* and *constructivism.*

I propose an intermediate position, which is motivated by the observation that the validity and applicability of quantum theoretical figures like complementarity and entanglement is not confined to quantum physics.[5] Although this is not the place to describe this position in detail[6], a few remarks should suffice to give an idea.

The world is never given to us directly and immediately, but always as an observed world and as it appears on our internal stage, i.e. inside us through our cognitive frame of reference. Every act of cognition is cognition of something, the observed, by someone, the observer. The *"epistemic cut"* separating the observed from the observer may be movable but it is never removable.

The modes under which the world appears to an observer are structured by *observables*. Observables correspond to possible questions or investigations the observer may ask or perform. Every act of cognition amounts to a *measurement* of an observable. That is, in a measurement the investigation related to the observable is really executed and yields a result, which can claim factual validity. The following fundamental features of observables and measurement are universally valid:

- A measurement can change the state of the observed system. Accordingly, a measurement will in general be more than a simple registration of a pre-given fact, but will to some extent create a fact.
- *Complementarity*: since measurements may change states, measuring an observable may destroy the facticity of the results of previous measurements of different observables. Quite commonly, different observables are *complementary*: it turns out to be impossible to attribute factual values to all of them simultaneously. Many examples of such a situation, in particular outside physics, have been worked out in the references quoted above.

- The observer is free to choose the observable he wants to measure but he has little or no control on the result of his measurement. This *recalcitrance of nature* is a strong drawback for any extreme form of constructivism.
- In this framework, concept formation is formation of observables. Observables exist neither entirely "out there" in the observed world nor completely inside the observer. They should be located astride on the epistemic cut. They are neither just detected nor mere inventions. Symmetries are a special case. One should attribute the ontological status of observables to them. This is the said intermediate position between essentialism and constructivism.
- There is much creative freedom in the subtle process of observable formation. The conceptualisation of the world is largely subjective and strongly dependent on the given cultural and historical context. However, this freedom is not unlimited. This is another instance of the recalcitrance of the world, which will reject unfitting conceptualisations.

## Virtuality, Freedom and Aesthetics

The contemplator of a symmetric object like a flower or a poem will be more or less aware of the underlying symmetry substitutions, but in general he will not really execute them. Symmetry substitutions reside in a status of virtuality rather than performed action. They are at home in the wide space of possibilities, the characteristic arena of existence of human beings, who, more than all other animals, are endowed with the capacity for exploration, conditionality and contrafactuality. It is the precondition of human freedom that we are not chained to blunt facts but are able to move in the infinite space of possibilities. Symmetries, too, are conceived as possibilities in various respects. Firstly, as already mentioned, symmetry operations remain in the status of possibilities. Secondly, the concrete realisation of a symmetry is in almost all cases only an approximate one. Perfect symmetries are idealisations residing in the realm of possibilities. Thirdly, an approximate realisation of a symmetry is just one of an infinity of possibilities, which are perceivable only under the vision of perfect symmetry.

Our considerations suggest that the notions of symmetry, order and freedom belong to a common conceptual basis in which also aesthetic considerations have their place. In fact, most people would agree that beauty is based upon a subtle interplay between order and freedom. The Greek word συμμετρία (symmetria) means "harmony" and clearly exhibits the strong aesthetic connotation of symmetry. Friedrich Schiller, who was not only a great poet but also a leading theoretician of aesthetics defines beauty as "*freedom in appearance*" [7]. The elements of a beautiful object arrange themselves in a common structure, which could also be quite different but, as it is, shows the signs of perfection. Beauty in this sense is unenforced and incompatible with rigidity, coercion, schematization or ideology.

This is also true for symmetry itself. Perfectly strict unbroken symmetry is normally felt to be too rigid and too much of a constraint to be beautiful. This is quite evident in the common attitude towards seriality and repetition. There is a widespread feeling that more than three repetitions are an insult and a violation of human dignity. Only up to three admonitions are common practice in legal acts and systems. In auctions, the number three plays a prominent ritualised role. On the other hand, symmetrical rigor, monotony and endless repetition can have their own fascination beyond beauty. We shall soon come back to this strange phenomenon.

"Freedom in appearance" requires a delicate balance between two conflicting desiderata: enough symmetry to maintain the visibility of an underlying structure and enough deviation from symmetry to give way to freedom and to avoid coercion and monotony. The equilibrium between symmetry and symmetry breaking is a compromise between the fulfilment of expectation and surprise, between harmony and dissonance. In works of art, this conflict may show up in the form of a relaxed and hilarious play or else as a dramatic struggle. There will always be an antagonism between "Apollonian" and "Dionysian" temperaments. Ecstatic insurgences against harmonising temperance will never end.

We already underlined the context dependence of symmetries and their role as an important special case of concept formation. Symmetries und serial variations in human artefacts have an effect on our level of consciousness. They interrupt the daily routine in which concepts are normally used, make the subtle process of concept formation transparent and refresh the awareness of our ability for the perception of similarities.

"Freedom in appearance" needs a sufficiently ample space of possibilities and, hence, a reasonable degree of complexity. From this point of view, minimalistic artistic productions like a black square are problematic. In such a case, the necessary complexity cannot lie within the artefact itself, but it may reside in the perplexity of the contemplator confronted with it.

## Symmetry, Life and Death

The delicate balance between symmetry and symmetry breaking, order and freedom, whose crucial importance in aesthetics we just discussed, is also a constitutive feature of the phenomenon of life. Life is a subtle, highly complex order structure, which is only able to maintain itself because it enjoys enough flexibility to react and actively adapt itself to its environment. Strict symmetry is incompatible with liveliness. This may be one reason why conscious living beings perceive perfect symmetry not so much as unresented beauty but rather with a feeling of discomfort and anxiety.

Before continuing, though, we have to deal with an obvious objection. Mathematicians and theoretical physicists invariably praise the beauty of mathematical structures. They may feel a certain thrill in connection to them, but the impression of beauty by far prevails. This is in part simply a matter of taste. They enjoy what frightens other people. But there are deeper reasons. Firstly, there is the pleasure of cleanness and transparency. Here, beauty lies in the absence of all the ugly, dirty, confused and trivial elements embittering our daily reality. Secondly and even more importantly, for mathematically gifted people mathematical structures are not at all dead but full of life and vibrating with symbolic meaning. The freedom of appearance, necessary for an aesthetic impression, is felt to be present to the highest possible degree.

A crystal is often looked at as a paradigm of lifelessness par excellence. There is, however, also a certain fascination inherent in such simple symmetric structures, which, for example, also seems to emanate from the Egyptian pyramids, though being clear monuments of death.

Similar is our ambivalent fascination with long or even infinite repetition. On the one hand, as already mentioned above, there is something insulting in their monotony. On the other hand, voluntary, self-imposed endless repetition can have a ruminant element like meditation. Eternity may flash up in it. A yearning for rest may lie in the fascination with repetition.

Sensitive persons may feel something frightening, lurking behind balanced beauty. The symmetry of a beautiful object is mitigated by slight deviation and freedom. But behind this the cold and lifeless rigidity of a perfect and unbroken symmetry may emerge as a vision. At some point, extreme beauty may turn into horror. In the words of R. M. Rilke's first elegy: [8]

*For beauty is only a step removed from a burning terror we barely sustain, and we worship it for the graceful sublimity with which it disdains to consume us.*

Something of this kind may also underlie the uneasy feeling about the Mandelstam set.

One need not adopt the Freudian concept of a parity of life and death drives to acknowledge the deep ambivalence and mutual dependence inherent in the relationships between symmetry and symmetry breaking, order and freedom and also death and life.

Rhythmic repetition in the form of the heart beat is essential for our feeling to be alive and yet it contains an element of deadly monotony. Rhythmic dance is a feast of life, but may also be seen as a dance of death. Both aspects are intimately entangled and it is impossible to decide which one prevails.

## About the Author

Hartmann Römer (*1943) held the position of a full professor for Theoretical Physics at the university of Freiburg, Germany, from 1979 up to his retirement in 2008. His research interests include quantum field theory, physics of elementary particles and mathematical physics as well as natural philosophy.